\documentclass{article}

\usepackage{microtype}
\usepackage{graphicx}
\usepackage{subfigure}
\usepackage{booktabs} 
\usepackage{pifont}
\usepackage{subcaption}  
\usepackage{multirow}

\usepackage{hyperref}
\usepackage{balance}

\usepackage[accepted]{icml2025}

\usepackage{amsmath}
\usepackage{amssymb}
\usepackage{mathtools}
\usepackage{amsthm}

\usepackage[capitalize,noabbrev]{cleveref}

\theoremstyle{plain}

\theoremstyle{definition}

\theoremstyle{remark}

\usepackage[textsize=tiny]{todonotes}

\newcommand\todoyang[1]{\textcolor{black}{#1}}

\newcommand{\xmark}{\ding{55}}
\newcommand{\cmark}{\ding{51}}

\newcommand{\squishlist}{
	\begin{list}{$\bullet$}
		{  \setlength{\leftmargin}{+0.15in}
		} }

		\newcommand{\squishend}{
	\end{list} }

\begin{document}

\twocolumn[
\icmltitle{Spatial Audio Processing with Large Language Model on Wearable Devices}




\icmlsetsymbol{equal}{*}

\begin{icmlauthorlist}
\icmlauthor{Ayushi Mishra}{equal,yyy}
\icmlauthor{Yang Bai}{equal,yyy}
\icmlauthor{Priyadarshan Narayanasamy}{yyy}
\icmlauthor{Nakul Garg}{yyy}
\icmlauthor{Nirupam Roy}{yyy}
\end{icmlauthorlist}

\icmlaffiliation{yyy}{Department of Computer Science, University of Maryland, College Park, USA}

\icmlcorrespondingauthor{Nirupam Roy}{niruroy@umd.edu}
\icmlcorrespondingauthor{Ayushi Mishra}{amishr13@umd.edu}
\icmlcorrespondingauthor{Yang Bai}{yangbai8@umd.edu}

\icmlkeywords{Machine Learning, ICML}

\vskip 0.3in
]



\printAffiliationsAndNotice{\icmlEqualContribution}

\begin{abstract}
Integrating spatial context into large language models (LLMs) has the potential to revolutionize human-computer interaction, particularly in wearable devices. In this work, we present a novel system architecture that incorporates spatial speech understanding into LLMs, enabling contextually aware and adaptive applications for wearable technologies. Our approach leverages microstructure-based spatial sensing to extract precise Direction of Arrival (DoA) information using a monaural microphone. 
To address the lack of existing dataset for microstructure-assisted speech recordings, we synthetically create a dataset called \textit{OmniTalk} by using the LibriSpeech dataset. 
This spatial information is fused with linguistic embeddings from OpenAI’s Whisper model, allowing each modality to learn complementary contextual representations. The fused embeddings are aligned with the input space of LLaMA-3.2 3B model and fine-tuned with lightweight adaptation technique LoRA to optimize for on-device processing. 
SING supports spatially-aware automatic speech recognition (ASR), achieving a mean error of 25.72°—a substantial improvement compared to the 88.52° median error in existing work—with a word error rate (WER) of 5.3.
SING also supports soundscaping, for example, inference how many people were talking and their directions, with up to 5 people and a median DoA error of 16°.
Our system demonstrates superior performance in spatial speech understanding while addressing the challenges of power efficiency, privacy, and hardware constraints, paving the way for advanced applications in augmented reality, accessibility, and immersive experiences.

\end{abstract}

\vspace{-0.3in}
\section{Introduction}
$\blacksquare$ {\bf Vision.}
Large language models (LLMs) have created new possibilities by enabling intuitive, contextual, and natural interactions with machine~\cite{choipicle, cui2024drive, kumaran2023scenecraft, chen2024magdi, alayrac2022flamingo}.
Introduction of this capability to smart earbuds and other wearables is around the corner~\cite{LLMonEarbuds}.
Spatial understanding of speech in LLMs can open up a frontier for wearable applications, as it enables such ubiquitous devices to process directional cues, recognize speech from multiple sources, and associate spoken content to its spatial context.
This capability can be a cornerstone in significantly enhancing user experiences by making devices more responsive, intuitive, and context-aware, particularly in applications like virtual assistants, augmented reality (AR), and accessibility tools. By enabling LLMs to reason about spatial acoustic cues, wearables can offer advanced features such as selective voice summarization in meetings, sound-based navigation for visually impaired individuals, and immersive XR experiences that react dynamically to the user’s surroundings.

Spatial context is yet to be explored to its full potential with LLMs.
BAT~\cite{zheng2024bat} is one of the first to support spatial understanding for LLM.
While inspiring, BAT supports inference only on non-verbal audio, such as dog barking or bird chirping, without the context of spoken language.
In this paper, we aim to advance the idea of introducing spatial knowledge in LLMs for speech signals and at the same time make it more compatible for wearables and ubiquitous computing devices.
This work strive to achieve high accuracy of directional audio sensing using a novel monaural setup and extending its capability to spatial speech processing.

$\blacksquare$ {\bf Challenges and Our Approach. }\\
While promising, realization of this vision needs to overcome several unique challenges.

{\bf A. Spatial cues in wearables. }
Sensing spatial features of sound, such as direction-of-arrival (DoA) or source location, requires sampling the wave in space using an array of microphones.
However, traditional methods are limited by space-time sampling constraints and requires large and power hungry microphone arrays~\cite{yang2022enhancing, wang2023fn} to achieve resolution of spatial cues essential for understanding acoustic physical contexts. 
These arrays require significant physical space, making them impractical for any miniature wearables and ubiquitous devices such as microscale IoT sensors or button-sized wearables.
BAT~\cite{zheng2024bat} achieves spatial understanding using monaural or binaural microphones. While claimed monaural, the mean error rate (MAE) is 88.52°, too high for real-world applications. 
The physical configuration of the microphones for their binaural setup was not mentioned.
As a solution to minimize the size of the setup for wearables, we break away from the traditional spatio-temporal sampling model and build on pioneering idea \cite{garg2021owlet} of micro-structure assisted miniature and low-power acoustic front-end.
This setup requires only a monaural microphone combined with a tiny microstructure to induce spatial diversity in the recorded signal, eliminating the need for multiple microphones while still enabling precise spatial encoding, making it an ultra-compact and low-power solution suitable for wearable applications.

{\bf B. Spatial speech context in LLM. } 
To capture critical spatial audio cues such as directionality and reverberation, we leverage the microstructure-based spatial sensing~\cite{garg2021owlet}, which enables \todoyang{compact setup for wearables and higher DoA estimation accuracy}. These cues are processed into embeddings using a lightweight architecture \todoyang{that only has 16.3 million parameters}.
The speech encoder utilized in this work is OpenAI's Whisper model~\cite{radford2023robust}.  
Its encoder generates features that are well-suited for modeling speech and effectively incorporate contextual information about background noises~\cite{gong2023whisper}.

{\bf C. Aligning spatial speech context to LLM. } 
To integrate spatial audio understanding into LLMs, we propose a two-step alignment framework that bridges the spatial encoder, Whisper's speech-to-text encoder, and the LLM. First, the spatial encoder processes acoustic signals to extract directional and spatial cues, such as DoA, and generates embeddings that encapsulate spatial information. Simultaneously, Whisper encodes speech signals into rich linguistic representations, capturing both semantic content and background context. 
We adopt the approach outlined in LLava~\cite{liu2024visual}, employing a simple linear layer as a projection matrix, denoted as \textit{W}, to map the spatial features into the language embedding space of the LLM. 
Then, the fused embeddings are aligned with the input space of the LLM using a lightweight adapter module and fine-tuned on task-specific prompts using low-rank adaptation (LoRA)~\cite{hu2021lora}. This approach allows the LLM to interpret spatially enriched embeddings and produce outputs that are both contextually and spatially aware, enabling advanced applications in spatial speech understanding and wearable interactions.

Our key contributions are summarized below:
\vspace{-0.1in}

\squishlist
    \item We present a novel framework that integrates spatial audio cues with LLMs, enabling advanced spatial speech understanding for real-world wearable applications.
    \vspace{-0.05in}
    
    \item We design a novel spatial encoder tailored for microstructure-based spatial sensing, enabling precise DoA estimation with minimal hardware. This compact and efficient encoder is specifically optimized for small, wearable devices.
    \vspace{-0.05in}

    \item Our framework enables wearable devices to perform advanced tasks such as spatially aware automatic speech recognition (ASR), meeting summarization with spatial context, and immersive audio experiences, showcasing the potential for applications in augmented reality, accessibility, and beyond.
    
\squishend

\vspace{-0.1in}
\section{Related Work}
\subsection{Spatial Audio Detection}
\vspace{-0.1in}
Spatial audio processing involves techniques from both traditional signal processing and modern deep learning approaches to extract spatial information from sound sources. Signal processing methods, such as beamforming ~\cite{xu2017waveforming}, generalized cross-correlation (GCC)~\cite{knapp1976generalized}, and eigenvector-based techniques like MUSIC~\cite{schmidt1986multiple} and ESPRIT~\cite{roy1989esprit}, utilize phase differences and time delays between microphone pairs to estimate the DoA. While effective, these methods often require large microphone arrays and can struggle in reverberant environments. Deep learning models, including convolutional neural networks (CNNs)~\cite{adavanne2018sound}, recurrent neural networks (RNNs)~\cite{xu2017convolutional}, and transformer-based architectures~\cite{park2021many}, have demonstrated success in learning spatial features from raw audio and spectrograms, often achieving superior results in complex scenarios, but microphone arrays are still required. Recent developments in microstructure-based sensing, where a monaural microphone is embedded in a designed physical structure to create directionally-dependent frequency responses, offer a promising approach for minimalist spatial audio sensing~\cite{garg2021owlet}. This method, inspired by biological systems like owl hearing, introduce new opportunities for low-cost, efficient DoA estimation while balancing physical constraints with data-driven learning.

\subsection{Multimodal Large Language Models}
\vspace{-0.1in}
Recent advancements in multimodal large language models (LLMs)\cite{yin2023survey, song2023bridge, huang2024large, hsieh2024sugarcrepe} have significantly expanded their capabilities across various modalities, including audio, music, and visual data. AudioGPT~\cite{huang2024audiogpt} integrates ChatGPT as a versatile interface for a wide array of audio and speech-related tasks, enabling complex reasoning and synthesis using natural language prompts. For music understanding, another work introduced a framework combining the MERT music encoder~\cite{li2023mert} with an LLM, achieving state-of-the-art results in tasks such as music captioning and mood classification.
While audio-based LLMs are gaining traction, multimodal LLMs have been more extensively explored in the visual domain~\cite{liimproving, li2023blip, liu2024visual, sun2024video}. Several models focus on image understanding by combining LLMs with advanced vision encoders. BLIP-2 ~\cite{li2023blip} employs a pre-trained Vision Transformer (ViT) to extract visual embeddings, which are then aligned with an LLM for tasks such as image captioning, visual question answering (VQA), and reasoning about visual content. LLaVA~\cite{liu2024visual} extends this framework with a lightweight visual encoder and cross-modal training strategies, enabling improved performance in open-ended visual reasoning tasks.
These advancements underscore the expanding landscape of multimodal LLMs, highlighting the importance of robust cross-modal architectures for diverse real-world applications across audio, music, image, and video tasks.

\subsection{Spatial-Aware Large Language Models}
\vspace{-0.1in}
LLMs have become increasingly popular and useful for various AI-driven applications~\cite{openai2023gpt4, touvron2023llama, vicuna2023, anthropic2023claude, mpt2023}. However, the integration of LLMs with physical context, particularly in spatial audio perception, remains sparse.
Recent papers, such as BAT~\cite{zheng2024bat} and {\em ``Can Large Language Models Understand Spatial Audio?''}~\cite{tang2024can}, have explored integrating spatial audio with LLMs, focusing on non-speech sounds and requiring larger microphone arrays. In contrast, our work targets spatial speech processing using monaural setup with microstructure sensing, enabling superior angular resolution with a smaller hardware footprint suitable for wearables. The goal for using the spatial speech is to enhance speech clarity and intelligibility by allowing listeners to distinguish multiple speakers based on spatial positioning. SALMONN~\cite{tang2023salmonn} and GAMA~\cite{ghosh2024gama} address speech tasks but lack spatial awareness, limiting their applicability to speech understanding. Our work bridges this gap, emphasizing low-power, privacy-preserving processing tailored for real-world wearable deployment.

\section{Microstructure-Assisted Spatial Encoding}
\subsection{Spatial Encoding Microstructure}
\begin{figure}[htbp]
\begin{center}
\includegraphics[width=3.3in]{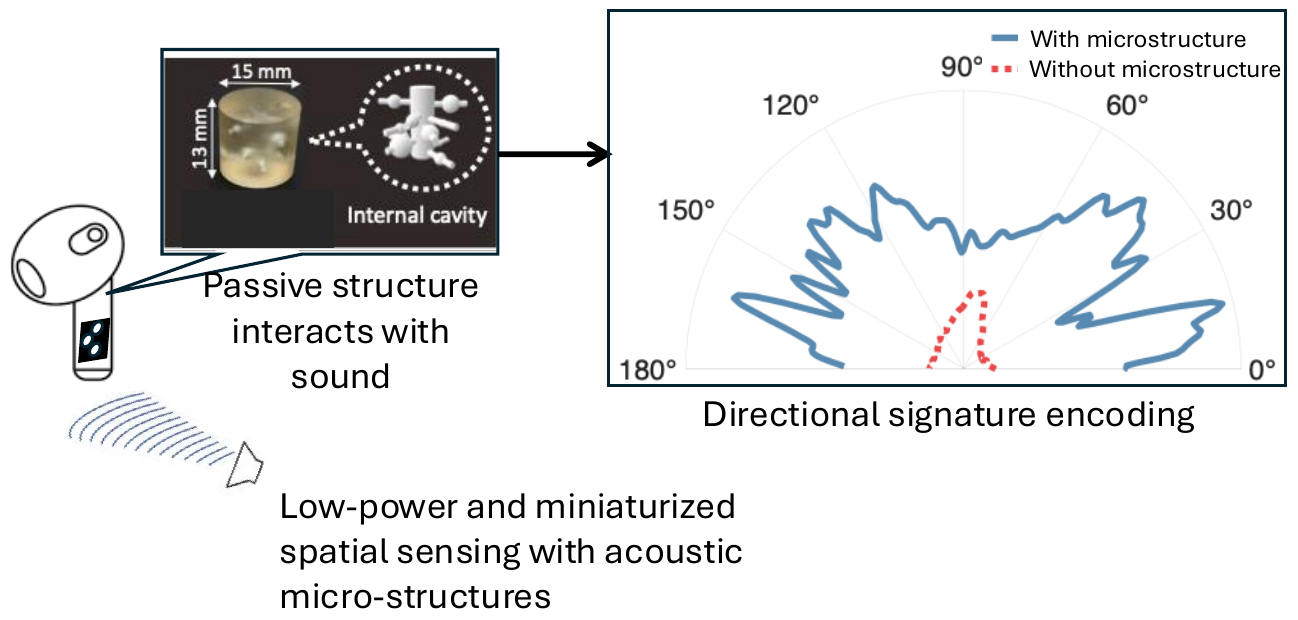}
\vspace{-0.3in}
\caption{\small{The vision and technical overview of Owlet~\cite{garg2021owlet}, a low-power and miniaturized system for introducing spatial information into monaural recording of sound. }}
\vspace{-0.1in}
\label{fig:owlet}
\end{center}
\end{figure}

Microstructure-assisted spatial encoding builds upon the innovative hardware design introduced in Owlet~\cite{garg2021owlet}, which utilizes a compact two-microphone array integrated with a microstructure to achieve high spatial resolution. As shown in Figure~\ref{fig:owlet}, the microstructure modifies sound wavefronts as they propagate, introducing distinct acoustic patterns that encode angular information. 
The Owlet system achieves spatial diversity in sound sensing with a monaural setup by leveraging the physical principles of diffraction, capillary effects, and structural resonance within a carefully designed microstructure. Diffraction occurs as sound waves bend and scatter when interacting with the edges and barriers inside the structure, creating phase shifts and amplitude variations that depend on the sound's DoA. Capillary effects arise from the narrow channels within the microstructure, where confined spaces alter sound propagation by introducing pressure gradients and phase delays, further emphasizing directional differences in the received signal. Additionally, structural resonance occurs when specific sound frequencies align with the natural resonance modes of the structure, amplifying certain frequency components while attenuating others based on the sound's angle of incidence. By combining these three mechanisms, the Owlet design introduces directionally dependent frequency responses, enabling precise DoA estimation with minimal hardware complexity, avoiding the need for large microphone arrays while maintaining high spatial resolution.

{\bf Monaural Recording Hardware:} In its original design, Owlet used two microphones, with an additional one placed outside the microstructure cap.  
This extra microphone helped reduce the impact of room impulse response on directional cues.  
Despite utilizing multiple microphones, the system produces a monaural recording, where the sound wave is essentially sampled at a single spatial location.  
Unlike binaural recording hardware, which uses two separated microphones to achieve spatial diversity, Owlet's hardware captures multiple recordings using closely placed microphones.  
As elaborated in Section~\ref{sec:spcontext}, we leveraged the second channel from the microphone outside the microstructure for speech encoding and the microstructure-covered microphone channel for DoA encoding.


\subsection{Spatial Speech Generation}
\vspace{-0.1in}
Capturing and annotating spatial speech in real-world environments is often a labor-intensive process, further complicated by variations in acoustic properties and the constraints of recording equipment. To efficiently generate a diverse dataset that covers a wide range of sound sources while ensuring comprehensive ground-truth metadata, a simulation-based approach has been adopted.

For this study, we utilized the Librispeech dataset~\cite{panayotov2015librispeech}, a publicly available corpus of high-quality English speech, sampled at 16KHz. The dataset provides phonetically diverse speech recordings, large vocabulary, and clean and noisy subsets, making it ideal for speech recognition and spatial analysis under different acoustic conditions~\cite{waywithwords2025speechdata}.
%
For the application of spatial ASR, we generated a comprehensive 400-hour spatial speech dataset. 
We pick 500 original samples from the LibriSpeech dataset, and convoluted them with the impulse responses from 1° to 360° with 1° resolution.

For the application of soundscaping, leveraging LibriSpeech, we generated a comprehensive 2,000-hour spatial speech dataset that simulates scenarios involving 1 to 5 speakers speaking simultaneously. 
To ensure robust and representative data, the number of speakers is evenly distributed across the dataset, with uniform coverage of DoA angles. This design enhances the dataset's robustness to variations in DoA, speaker count, and speaker diversity. Multi-speaker samples were created by randomly selecting speech segments and assigning distinct DoA values, ensuring realistic spatial distributions and promoting generalization to real-world scenarios. Our dataset, named \textbf{\textit{OmniTalk}}, symbolizes the 360° spatial diversity and multi-speaker dynamics it embodies, offering a robust foundation for spatial speech processing.

\textbf{Spatial Speech Generation Process}: We created a synthetic spatial speech dataset by convolving the clean Librispeech signals with Owlet-specific impulse response at discrete angles. The system utilizes a set of frequency-domain impulse response \( H(\omega, \theta) \), where \( \omega \) represents the angular frequency and \( \theta \) denotes the angle of arrival (in degrees). 
Since the microstructure has a consistent shape, we did a one-time calibration of the impulse response. 
These responses represent how the Owlet microstructure-based array processes incoming speech signals from different directions. 
To convert the frequency-domain representation into the time domain, an \textit{Inverse Fast Fourier Transform (IFFT)} is applied:

\begin{equation}
h_\theta(t) = \mathcal{F}^{-1}\{H(\omega, \theta)\}
\end{equation}

where \( h_\theta(t) \) is the time-domain impulse response for angle \( \theta \). This transformation is performed for all the desired angles \( \theta \in [1, 360] \), yielding angle-specific impulse responses. Each speech file \( y_{\text{original}}(n) \), sampled at original rate \( f_{\text{original}} \) is loaded and resampled to a uniform sampling frequency \( f_s = 16\,\text{kHz} \). This ensures consistency across all signals for accurate convolution wiht the impulse responses. The resampled signal \( y(n) \) is obtained as:

\begin{equation}
y(n) = \text{Resample}\big(y_{\text{original}}(n), f_s, f_{\text{original}}\big)
\end{equation}

where the resampling operation adjusts the temporal resolution of the signal while preserving it's content. 

To simulate how speech signal is perceived at angle \( \theta \), the resampled speech signal \( y(n) \) is convolved with the corresponding impulse response \( h_\theta(n) \). The discrete-time convolution operation is defined as:
\begin{equation}
y_{\text{conv}, \theta}(n) = (y * h_\theta)(n) = \sum_{m=-\infty}^{\infty} y(m) \cdot h_\theta(n - m)
\end{equation}
where, 
\vspace{-0.1in}
\squishlist
    \item \( y(m) \) is amplitude of the speech signal at time index \( m \).
    \vspace{-0.05in}
    \item \( h_\theta(n - m) \) is the impulse response of the system for angle \( \theta \), delayed by \( m \) samples.
    \vspace{-0.05in}
    \item \( y_{\text{conv}, \theta}(n) \) is the resulting signal after the speech is filtered by the spatial characteristics at angle \( \theta \).
\squishend

Refer to the spectrogram of recordings from different directions in Appendix~\ref{app:spectrogram}.

Our dataset provides a high-fidelity synthetic spatial speech corpus tailored for evaluating monaural spatial sensing models. By leveraging Owlet-specific impulse responses, our dataset accurately simulates how speech signals are perceived at different angles, ensuring realistic spatial encoding. The one-time calibrated impulse responses maintain consistency across experiments, eliminating variations caused by environmental changes. Furthermore, the dataset is derived from Librispeech, a widely used speech corpus, ensuring high-quality linguistic content. The rigorous preprocessing steps, including uniform resampling and frequency-to-time domain transformation via IFFT, guarantee temporal alignment and signal integrity. By applying discrete-time convolution with angle-specific impulse responses, our dataset faithfully replicates the directional filtering imposed by the Owlet microstructure, making it an ideal benchmark for spatial speech recognition and DoA estimation tasks.

\begin{figure*}[htbp]
\begin{center}
\includegraphics[width=\textwidth, keepaspectratio]{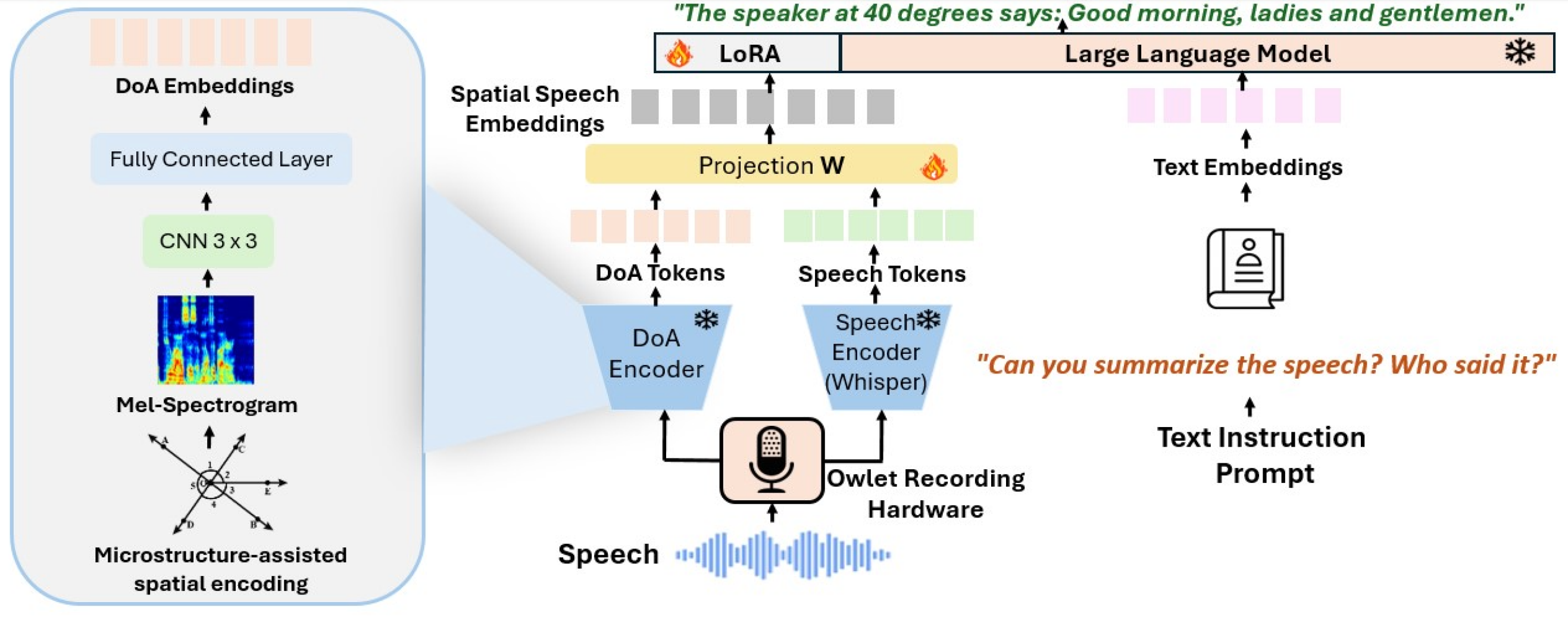}
\vspace{-0.30in}
\caption{\small{Spatial-aware framework for direction and speech transcription.}}
\vspace{-0.20in}
\label{fig:model_Structure}
\end{center}
\end{figure*}

\vspace{-0.15in}
\section{SING: Spatial Context to Wearable LLM}
\label{sec:spcontext}
Our system is designed to integrate spatial speech understanding into LLMs while addressing the unique constraints of wearable devices. It consists of three main components: spatial speech sensing, embedding alignment, and LLM fine-tuning. As shown in Figure~\ref{fig:model_Structure}, the spatial speech sensing module uses a Owlet recording hardware to capture high-resolution spatial cues, such as directionality, without the need for a traditional large microphone array. 
The recordings from the two channels of the Owlet are the input for the DoA encoder and the speech encoder. 
Spatial cues are processed into low-dimensional embeddings using lightweight models. Whisper encodes raw speech into dense speech embeddings that capture linguistic features. These embeddings together with spatial embeddings are then aligned with the input space of LLMs. The LLM is then fine-tuned on our spatial speech dataset, enabling it to interpret speech and DoA. 





\subsection{Spatial Speech Encoder}\label{sec:spatial_speech_encoder}
\textbf{Speech signal preprocessing.}
Different from existing work that uses an array of microphones for spatial sensing, our system uses only a monaural microphone inside the microstructure. 
Given the different impulse responses from directions of speech, the amplitude of the signal changes overtime, and different frequencies have different patterns of amplitude change. 
Due to this nature, we convert the captured signal from time domain $x(n)$ to frequency domain $X(t, f)$ using STFT
\begin{equation}
   X(t, f) = \sum_{n=0}^{N-1} x(n) \cdot w[n - t] \cdot e^{-j\frac{2\pi fn}{N}} 
\end{equation}
where w(n) represents a Window function of length N. Then we convert the output into mel scale, 
\begin{equation}
S_{\text{mel}}(t, j) = \log \left( \sum_{k=1}^{K} |X(t, f)|^2 \cdot H_j(k) + \epsilon \right)
\end{equation}
where $H_j(k)$ is Mel filter for the j-th band, and $\epsilon$ is a small constant to avoid log of zero.


\textbf{DoA Encoder.} 
The DoA encoder is a convolutional neural network, balancing computational efficiency with robust feature extraction. The architecture consists of three sequential convolutional blocks, each comprising a 2D convolutional layer, batch normalization for stabilizing training, a ReLU activation function, and max-pooling for spatial dimension reduction. A flattening layer reshapes the output from the convolutional layers into a 1D vector, which is then passed through a fully connected layer with 512 units, followed by a dropout layer to mitigate overfitting. The final linear layer maps the high-dimensional feature vector to DoA prediction output.
Refer to the 3D UMAP visualization of the embeddings from different angles in Appendix~\ref{app:umap}.

\subsection{Speech-to-text Encoder.}
\vspace{-0.1in}
The speech encoder utilized in this work is OpenAI's Whisper model~\cite{radford2023robust}.  
Its encoder generates features that are well-suited for modeling speech and effectively incorporating contextual information about background noises~\cite{gong2023whisper}. To extract Whisper embeddings, speech files are first processed to ensure consistency in sampling rate, mono channel format and fixed duration by padding to 30 seconds. The processed waveforms are passed through the \textit{WhsiperProcessor} for feature extraction, which generates mel-spectrogram features as input to the model. We use a whisper-large-v3\cite{huggingface_whisper} model for generating features. The encoder outputs hidden states for each time frame, representing semantic and acoustic information. To reduce the dimensionality while preserving critical information, adaptive average pooling is applied to the hidden states along the temporal axis, yielding fixed-size embeddings of shape [POOL\_SIZE, hidden\_dim] for each speech file, where POOL\_SIZE is 128 and hidden\_dim is 1024. The embeddings illustrate both temporal and semantic characteristics of the speech for downstream tasks such as transcription. 

\subsection{Alignment to LLM Space}\label{sec:llm_align}
\textbf{Pretraining.}
The spatial encoder in our system is pre-trained on a spatial ASR task consisting of two subtasks: DoA prediction, and speech recognition. The DoA task estimates the directional angles of each speaker.
The ASR task transcribes spoken speech into text. The two tasks are optimized using cross-entropy loss.
The spatial embeddings derived from DoA and the Whisper encoder are not inherently aligned with the input space of LLMs. To address this, we adopted the approach outlined in LLava~\cite{liu2024visual}, employing a simple linear layer as a projection matrix, denoted as \textit{W}, to map the spatial features into the language embedding space of the LLM. This linear projection layer was selected for its lightweight design, offering a more efficient and streamlined solution compared to the Q-Former connection module used in BLIP2~\cite{li2023blip}, the perceiver resampler and cross-attention layers in Flamingo~\cite{alayrac2022flamingo}. This design choice prioritizes efficiency and simplicity, ensuring seamless integration without adding significant complexity to the model architecture. 


\textbf{Supervised Fine-Tuning.}
We fine-tuned our pre-trained model using an instruction fine-tuning approach to enhance the LLM's ability to follow human instructions and provide greater control over its output. Specifically, we adopted a supervised fine-tuning (SFT) strategy~\cite{huggingface_sft_trainer}, enabling the model to learn from spatial embeddings, prompts, and their corresponding responses under direct supervision. For this purpose, we use the LLaMA 3.2 3B model~\cite{touvron2023llama} with LoRA~\cite{hu2021lora}, a method that optimizes the fine-tuning process for spatial speech understanding. (Refer to the detailed prompts in Appendix~\ref{app:prompts}) Instead of modifying the entire model, LoRA introduces trainable low-rank matrices into the lower layers of the LLM, significantly reducing computational and memory overhead while accelerating training. 

\vspace{-0.15in}
\section{Soundscaping: Multi-DoA Encoder}
Soundscaping for LLM will enhance human-AI interaction by creating more natural, immersive and context-aware auditory experiences.  We present the first approach to incorporate multi-direction of arrival (multi-DoA) information into an LLM. Our multi-DoA encoder consists of two key components: a number-of-speaker encoder and a DoA encoder. The number-of-speaker encoder predicts the count of active speakers in the environment. (Refer to the performance of number-of-speaker encoder in Appendix~\ref{app:num-of-speaker-encoder}) 
Based on the number of speakers, we call the corresponding DoA encoder, capturing the precise DoA angles.
We train 5 DoA encoders, ranging from 1 to 5 speakers.
The embeddings generated by the number-of-speaker encoder and the corresponding DoA encoder are concatenated to form a unified representation, which serves as the output of our spatial speech encoder $Z$ as:
\begin{equation}
    Z_{spatial} = Concat(Num-speaker(X), DoA(X)),
\end{equation}
where $X$ is a 8-second speech input, Num-speaker(·) and DoA(·) are
the number-of-speaker encoder and DoA enoder, Concat(·) is the concatenation operation along the feature dimension.
The number-of-speaker encoder follows same structure as the DoA encoder introduced in Section~\ref{sec:spatial_speech_encoder}.
The DoA encoders share the same structure involves adjusting the final fully connected layer to produce a regression output representing multiple DoAs sorted in ascending order. 
This change allows the model to handle dynamic outputs while maintaining the original convolutional structure for feature extraction.
The concatenated embedding is also aligned to the LLM space following the same strategy as described in Section~\ref{sec:llm_align}. We prepared question answer pair for finetuning the LLM model generating contextual responses for multiple DoA detection. 

\vspace{-0.15in}
\section{Evaluation of Spatial-Aware ASR}
\subsection{Dataset} 
\vspace{-0.1in}
To evaluate the proposed microstructure-based spatial speech encoder, there are no publicly available real or synthetic datasets that consist of general speech. 
Since DNN-based methods need sufficiently large
datasets to train on, we first calibrate the impulse response of the microstructure from all angles, and then create a large synthetic dataset by convolving the calibrated impulse response with the speech samples in the LibriSpeech dataset, as described in Section 3.2. 
After convolution with the impulse responses, we
trim or pad these clips to 30 seconds. The resulting waveforms are monaural with only 1 channel at a 16kHz sampling rate. 
We use a window size of 400, a hop size of 160, and 80 mel-bins to compute the Short-Time Fourier Transforms (STFTs) and mel-spectrograms.
As a result, for a 30-second recording, the Mel-spectrogram feature dimension is (1, 80, 3000).


\subsection{Training Details}
\vspace{-0.07in}
\textbf{DoA encoder.} The encoder is trained on 1 A100 GPU, with each epoch taking approximately 20 minutes. 
The encoder was first trained as a regression task where the output are the DoA values.
We take the last fully-connected layer as the embedding.

\textbf{LLM alignment and fine-tuning.}
This part is separated into two stages. We first lock the LLM model to train parameters for the projection layer. 
After the loss has converged to stable, we fine-tune the LLM model using LoRA to enhance
the LLM’s ability to follow human instructions and provide greater control over its output.
The training is completed on 3 H100 80 GB GPUs. 
Refer to the hyperparameters for training in Appendix~\ref{app:hyperparameters}.

\subsection{Performance on Spatially Aware ASR}
\vspace{-0.1in}
\begin{table*}[htbp]
\centering
\caption{Comparison of Performance on Spatially Aware ASR with One Sound Source. BAT supports DoA estimation and audio source type recognition, while SALMONN focuses on speech ASR and other LLM functions but lacks spatial awareness. Our model uniquely integrates DoA and ASR for spatially aware ASR.}
\label{tab:spatial_asr_comparison}
\begin{tabular}{lll}
\toprule
\textbf{Model} & \textbf{Supported Task} & \textbf{Metric / Performance} \\
\midrule
BAT            & DoA Estimation + Audio Source Recognition (No Speech) & MAE (°) ↓: 88.52 \\
SING [our work]           & DoA + Speech ASR (Spatial Awareness)       & MAE (°) ↓: \textbf{25.72} \\
\midrule
SALMONN        & Speech ASR and LLM Functions (No Spatial Awareness) & WER (\%) ↓: 2.2 \\
SING [our work]            & Speech ASR (without DoA)                 & WER (\%) ↓: \textbf{1.8} \\
SING [our work]            & DoA + Speech ASR (Spatial Awareness)                 & WER (\%) ↓: 5.3 \\
\bottomrule
\end{tabular}
\vspace{-0.1in}
\end{table*}

There is currently no existing work that integrates both DoA estimation and ASR within a LLM framework. To provide a comprehensive evaluation, we separately compare our model's DoA performance with BAT and its ASR performance with SALMONN, as shown in Table 2.

Since we only use the monaural microphone inside the microstructure for DoA estimation, we compare our performance with BAT in monaural case. Our model achieves a significantly lower Mean Absolute Error (MAE) of 25.72° compared to BAT's 88.52°. This result underscores the strength of our microstructure-based spatial encoding, which enables precise localization of sound sources in a computationally efficient manner, outperforming BAT's approach.
Our system also shows robust performance under noise and room reverberation, as shown in Appendix~\ref{app:noise-room_reverberation}.
Moreover, we evaluated the performance for three other speech datasets and one audio dataset. Further details can be found in Appendix~\ref{app:datasets}.

In terms of ASR, while our model achieves a word error rate (WER) of 5.3\% for speech recognition in the spatial ASR configuration, which is higher than SALMONN's WER of 2.2\%, it is important to highlight that our model incorporates spatial features, a capability absent in SALMONN. Integrating spatial information, such as DoA, introduces additional complexity to the task by requiring the model to jointly process acoustic and spatial cues. This enables advanced spatial reasoning and makes our approach more versatile for applications involving spatially distributed sound sources. Furthermore, when spatial features are excluded, our model achieves a WER of 1.8\%, outperforming SALMONN and demonstrating its strong baseline performance for speech recognition. These results underscore that the inclusion of spatial features, while slightly increasing the WER, significantly broadens the utility of the model, offering a trade-off between basic recognition accuracy and enhanced spatial awareness.
Refer to the qualitative examples in Appendix~\ref{app:asr_responses}.


\section{Evaluation of Multi-DoA Encoder}
\subsection{Dataset}
\vspace{-0.1in}
Five datasets are generated with no temporally overlapping sources, maximum two overlapping sources, maximum three overlapping sound sources, four overlapping sound sources, and five overlapping sound sources.
We start synthesizing a recording by randomly choosing the number of speech samples. 
Then we randomly pick the number of DoAs that at least 10 degrees of separations are guaranteed between sound sources to avoid spatial overlapping. 
After convolving the speeches with the impulse response of its DoA, the convoluted speeches are summed together as the final recording.
The final recordings are first loudness normalized by scaling them so that each clip has the same total sound energy. 
The final recording is trim or pad to 8 seconds. The resulting waveforms are monaural with only 1 channel at a 16kHz sampling rate. 
We use a window size of 2048, a hop size of 512, and 128 mel-bins to compute the Short-Time Fourier Transforms (STFTs) and mel-spectrograms.
As a result, for a 8-second recording, the Mel-spectrogram feature dimension is (1, 128, 251).

\subsection{Baseline}
\vspace{-0.1in}
Since our approach enables DoA estimation using a monaural microphone, which is fundamentally not achievable with traditional signal processing methods like MUSIC~\cite{kundu1996modified, tang2014doa} that rely on multiple microphones, we instead compare our model against deep learning-based algorithms, as they can be designed to handle both monaural and array-based spatial sensing setups.
We compare our performance with array-based speech DoA detection SELDNet~\cite{adavanne2018sound}. Following the evaluation of BAT~\cite{zheng2024bat}, we train the monaural-microphone based model AudioMAE~\cite{huang2022masked} using our microstructure-encoded monaural-microphone dataset for DoA estimation. 

\begin{table*}[t]
\centering
\caption{Comparison of MAE DOA error and MEEM across models with a known number of active sources. MEEM is calculated as $\text{MSE} \times (\text{Number of Microphones})$.}
\label{tab:mae_neem_comparison}
\begin{tabular}{llcccccc}
\toprule
\textbf{Model} & \textbf{Metric} & \textbf{1 Source} & \textbf{2 Sources} & \textbf{3 Sources} & \textbf{4 Sources} & \textbf{5 Sources} \\
\midrule

\multirow{3}{*}{SELDNet} 
 & MAE ($\downarrow$)           & 90.03 & \xmark & \xmark & \xmark & \xmark \\
 & MEEM ($\downarrow$)          & 360.12 & \xmark & \xmark & \xmark & \xmark \\
 & Median Error ($\downarrow$)  & 90.14 & \xmark & \xmark & \xmark & \xmark \\

\midrule

\multirow{3}{*}{AudioMAE} 
 & MAE ($\downarrow$)           & 43.79 & \xmark & \xmark & \xmark & \xmark \\
 & MEEM ($\downarrow$)          & 43.79 & \xmark & \xmark & \xmark & \xmark \\
 & Median Error ($\downarrow$)  & 27.79 & \xmark & \xmark & \xmark & \xmark \\

\midrule

\multirow{3}{*}{SING (Ours)} 
 & MAE ($\downarrow$)           & \textbf{25.72} & \textbf{24.16} & \textbf{28.11} & \textbf{23.31} & \textbf{17.08} \\
 & MEEM ($\downarrow$)          & \textbf{25.72} & \textbf{24.16} & \textbf{28.11} & \textbf{23.31} & \textbf{17.08} \\
 & Median Error ($\downarrow$)  & \textbf{13.00} & \textbf{13.00} & \textbf{20.00} & \textbf{18.00} & \textbf{13.00} \\

\bottomrule
\end{tabular}
\vspace{-0.1in}
\end{table*}

\subsection{Evaluation Metrics} 
\vspace{-0.1in}
To comprehensively evaluate the performance of our model and facilitate fair comparisons with prior works, we employ three metrics: Mean Absolute Error (MAE),  the Modified Error Efficiency Metric (MEEM), and median error. MAE is a widely used metric that quantifies the average angular difference, in degrees, between the estimated and ground-truth DoA. This metric directly evaluates the accuracy of the DoA predictions, offering an intuitive understanding of the model’s spatial localization capabilities.
In addition to MAE, we introduce MEEM, which is designed to normalize the performance by accounting for the number of microphones used in the system. MEEM is calculated as MEEM=MSE×(Number of Microphones), where MSE represents the mean squared error of the DoA predictions. Unlike traditional metrics, MEEM provides a balanced view of model efficiency by penalizing setups that require a larger number of microphones, making it particularly useful for comparing models designed for minimalist setups like ours. By combining these two metrics, we not only assess the raw accuracy of the models but also highlight the efficiency of our approach in leveraging a monaural microphone compared to multi-microphone arrays. 
We also show the median error of DoA as a better understanding of the distribution of errors.

\subsection{Performance of DoA Estimation}
\vspace{-0.1in}
Table~\ref{tab:doa_error} presents the DoA estimation performance of various models under scenarios with a known number of active sources. 
SELDNet, utilizing four microphones, shows relatively high MAE error for a single source (90.03°).
Due to the lack of spatial information in AudioMAE, it shows a MAE of 43.79° and a median error of 27.79°.
Our proposed model, using only one microphone, achieves competitive performance despite its minimalist hardware setup. For one source, it yields an MAE of 25.72° and median error of 13.00°, which outperforms AudioMAE, showing the efficiency of microstructure. Notably, for scenarios involving two or more sources, our model demonstrates robust performance with MAE values of 24.16°, 28.11°, 23.31°, and 17.08° for two, three, four, and five sources, respectively. 
Refer to the CDF plots in Appendix~\ref{app:cdf}.
These results highlight the ability of our monaural-microphone approach to provide reasonable DoA estimation accuracy while minimizing hardware complexity, making it suitable for wearable and low-power applications. Additionally, the MEEM values underline the efficiency of our approach in leveraging minimal hardware without significant performance trade-offs.
For qualitative results, refer to Appendix~\ref{app:doa_response}.

\section{Discussion and Future Work}
\vspace{-0.1in}
Our work demonstrates the feasibility of integrating spatial speech understanding into LLMs through a novel combination of spatial audio encoders, DoA estimation, and speech recognition. 
By leveraging compact microstructure-based sensing, we enable accurate speech transcription and spatial awareness using significantly compact microphones compared to traditional systems. 
Our method embeds spatial cues directly within a monaural recording, allowing precise DoA estimation in an extremely compact design, unlocking new possibilities for minimalist perception approaches and wearable sensing technologies.
While our model achieves competitive performance in both numerical and qualitative evaluations, several limitations and opportunities for future exploration remain.


\textbf{Future work.} Future work will focus on extending the spatial encoder to support elevation angle estimation, enabling full 3D spatial audio processing. Furthermore, collecting real-world datasets with diverse acoustic conditions and speaker configurations will enhance the robustness and applicability of the system. Another promising direction is the integration of multimodal inputs, such as combining visual information from cameras with audio, to improve spatial understanding and enable richer context-aware applications. We also plan to optimize the system for low-power hardware, ensuring its viability for real-time processing on wearable devices. 


\section{Conclusion}
In this work, we introduced a framework for spatial speech understanding integrated with LLMs, enabling directional audio processing and speech recognition on compact wearable devices. Through a microstructure-based spatial encoder and alignment with Whisper embeddings, our system achieves DoA estimation and speech ASR performance with minimal hardware resources. This work paves the way for enhancing wearable technologies in applications such as augmented reality and accessibility, laying a foundation for future advancements in spatially aware LLMs.

\section{Acknowledgment}
This work was partially supported by NSF CAREER Award 2238433. We also thank the various companies that sponsor the iCoSMoS laboratory at UMD.

\section*{Impact Statement}
This work introduces a novel framework for integrating spatial audio context into LLMs, specifically designed for wearable devices. By leveraging microstructure-based spatial sensing and efficient fusion techniques, this research addresses key challenges in spatial speech processing, such as computational efficiency, privacy, and hardware constraints. The proposed system paves the way for transformative applications in augmented reality, accessibility, and interactive computing, enabling context-aware and intelligent interactions between humans and machines. Beyond its technical contributions, this work has the potential to improve accessibility for individuals with disabilities, enhance safety in search-and-rescue operations, and redefine user experiences in immersive environments. While promising, the ethical implications of deploying such systems, including privacy and responsible use, must be carefully considered to maximize societal benefits.

\nocite{langley00}
\bibliography{reference}
\balance
\bibliographystyle{icml2025}

\newpage
\appendix
\onecolumn
\section{Towards Spatially Aware Language Models}

Large Language Models (LLMs) have proven invaluable for a wide range of applications by assisting with complex tasks such as legal text interpretation and other specialized domains including healthcare, finance and academic research. However, as human interactions and real-world scenarios increasingly demand context that goes beyond textual or verbal cues, integrating spatial knowledge into LLMs has become essential. Humans not only understand semantic context but also perceive and interpret physical cues such as sound and location, which inform more holistic decision-making in daily life. In this vein, the physics-aware LLMs can power applications like autonomous robotics that need to navigate and interact with three-dimensional spaces, or digital-twins that can simulate real-world environments for urban planning, energy management, and supply chain optimization. By grounding LLMs in the realm of physical space, we ensure these models can provide more accurate, relevant and human like support, bridging the gap between abstract textual reasoning and tangible real-world contexts.

\section{Hyperparameters}\label{app:hyperparameters}

SING presents specific hyperparameters for DoA encoder, LLM pretraining and finetuning in Table ~\ref{tab:hyperparameters}.

\begin{table*}[ht]
\centering
\begin{tabular}{lccc}
\hline
\textbf{Hyperparam}        & \textbf{DoA Encoder} & \textbf{LLM Pretraining} & \textbf{LLM Fine-Tuning} \\
\hline
\textbf{batch\_size}        & 32                  & 8                       & 8                         \\
\textbf{num\_epochs}        & 20                  & 5                       & 10                        \\
\textbf{learning\_rate}     & 0.001               & 1e-5                    & 1e-5                      \\
\textbf{patience}           & 5                   & \xmark                  & \xmark                    \\
\textbf{warmup\_steps}      & \xmark              & 0                       & 0                         \\
\textbf{loss\_function}     & MSELoss             & Causal LM loss          & Causal LM loss            \\
\textbf{optimizer}          & Adam                & AdamW                   & AdamW                     \\
\textbf{scheduler}          & ReduceLROnPlateau   & Linear                  & Linear                    \\
\textbf{gradient\_chkpt}    & \xmark              & Enabled                 & Enabled                   \\
\textbf{max\_seq\_length}   & \xmark              & 512                     & 512                       \\
\textbf{LoRA\_rank (r)}     & \xmark              & \xmark                  & 8                         \\
\textbf{LoRA\_alpha}        & \xmark              & \xmark                  & 16                        \\
\textbf{LoRA\_dropout}      & \xmark              & \xmark                  & 0.1                       \\
\textbf{device}             & CPU/GPU             & Multi-GPU Dist.         & Multi-GPU Dist.           \\
\textbf{mixed\_precision}   & \xmark              & FP16 autocast           & FP16 autocast             \\
\hline
\end{tabular}
\caption{Key hyperparameters for the DoA Encoder, LLM Pretraining, and LLM Fine-Tuning. 
`\xmark` indicates that the hyperparameter does not apply.}
\label{tab:hyperparameters}
\end{table*}

\section{Responses Generated for Spatial-Aware ASR}\label{app:asr_responses}
We present several examples of the spatial transcriptions from SING as shown in Table ~\ref{tab:spatial_transcriptions}. 

\begin{table*}[h!]
\centering
\begin{tabular}{@{}p{\dimexpr\textwidth-2\tabcolsep}@{}}
\toprule
\textbf{Spatial Transcriptions}\\ \midrule
The speaker is speaking approximately at angle 61 degrees. The speech says:  Lectures. \\ 
The speaker is speaking approximately at angle 182 degrees. The speech says:  The Emperor's Daughter. \\ 
The speaker is speaking approximately at angle 140 degrees. The speech says:  The Wandering Singer. \\ 
The speaker is speaking approximately at angle 96 degrees. The speech says:  Tom nodded worriedly. \\ 
\bottomrule
\end{tabular}
\caption{Spatial-aware ASR.}
\label{tab:spatial_transcriptions}
\end{table*}

\section{Responses Generated for Multiple Direction of Arrival Detection}\label{app:doa_response}
Table~\ref{tab:response_table1} presents several examples of number of sound source inference and the DoA detections from SING.

\begin{table*}[h!]
\centering
\begin{tabular}{@{}p{2.5cm}p{\dimexpr\textwidth-3cm-2\tabcolsep}@{}}
\toprule
\textbf{Num of Sources} & \textbf{Response Generated} \\ \midrule
1 & There is one speech source. The speech source's degree of arrival is 77.0 degrees. \\ \midrule
2 & There are 2 speech sources. Speech source 1's degree of arrival is 39.0 degrees, and speech source 2's degree of arrival is 101.0 degrees. \\ \midrule
3 & There are 3 speech sources. speech source 1's degree of arrival is 18 degrees, speech source 2's degree of arrival is 39 degrees, and speech source 3's degree of arrival is 257 degrees. \\ \midrule
4 & There are 4 speech sources. Speech source 1's degree of arrival is 3 degrees, speech source 2's degree of arrival is 118 degrees, speech source 3's degree of arrival is 203 degrees, and speech source 4's degree of arrival is 242 degrees. \\ \midrule
5 & There are 5 speech sources. Speech source 1's degree of arrival is 34 degrees, speech source 2's degree of arrival is 161 degrees, speech source 3's degree of arrival is 195 degrees, speech source 4's degree of arrival is 234 degrees, and speech source 5's degree of arrival is 317 degrees. \\ 
\bottomrule
\end{tabular}
\caption{Response generated for multiple DoA detection.}
\label{tab:response_table1}
\end{table*}

\section{Performance under Noise and Room Reverberation}\label{app:noise-room_reverberation}

To evaluate the impact of noise and reverberation on our system's performance, we leverage the GTU-RIR dataset~\cite{pekmezci2024evaluation}. This dataset provides high-fidelity room impulse responses (RIRs) measured in diverse acoustic environments. including 11 rooms with types of classroom, conference hall, sports hall, office, hotel room and staircase, making it ideal for simulating realistic reverberation conditions in our evaluation. 
To systematically assess the robustness of our model, we conduct experiments under controlled conditions with varying levels of signal-to-noise ratio (SNR) and reverberation. Table~\ref{tab:noise_reverberation} presents the results, showing the Mean Absolute Error (MAE) and Median Error in degrees across different test conditions. Our findings indicate that while reverberation and noise impact DoA accuracy, our system remains robust compared to existing work. Specifically, without noise or reverberation, we achieve a MAE of 25.72°, which increases to 48.69° under reverberation, demonstrating the effect of multipath reflections. The addition of noise at different SNR levels further degrades performance, with errors increasing as noise levels rise.
These results underscore the importance of robust spatial encoding techniques in real-world applications, particularly in noisy and reverberant environments. Future work will focus on adaptive denoising strategies and spatial de-reverberation methods to enhance the resilience of our framework in challenging acoustic conditions.

\begin{table}[h]
    \centering
    \caption{Performance under noise and room reverberation: We employ our proposed architecture to infer under distinct scenarios, with and without noise (different SNR in dB levels) and and reverberation.}
    \label{tab:ablation}
    \begin{tabular}{cc cc}
        \toprule
        \textbf{Noise} & \textbf{Reverberation} & \textbf{MAE (°)} $\downarrow$ & \textbf{Median Error (°)} $\downarrow$ \\
        \midrule
        \xmark & \xmark & \textbf{25.72} & \textbf{13.00} \\
        \xmark & \cmark & 48.69 & 28.42 \\
        60dB   & \cmark & 48.57 & 28.27 \\
        50dB   & \cmark & 49.50 & 28.95 \\
        40dB   & \cmark & 50.89 & 31.08 \\
        30dB   & \xmark & 33.01 & 21.25 \\ 
        30dB   & \cmark & 54.43 & 33.12 \\
        \bottomrule
    \end{tabular}
    \label{tab:noise_reverberation}
\end{table}

\section{Performance on Different Datasets}\label{app:datasets}
To assess the generalizability of our proposed spatial-aware ASR system, we evaluate its performance across multiple datasets, including our synthetically generated dataset based on LibriSpeech, as well as other speech datasets Common Voice~\cite{ardila2019common}, VoxCeleb~\cite{nagrani2017voxceleb}, and LJ Speech~\cite{ljspeech2017}. Our results in Table~\ref{tab:dataset_performance} indicate that the model achieves the lowest MAE and Median Error, benefiting from the model trained using the same dataset. However, when tested on other datasets, the MAE and Median Error slightly increases as expected. Despite this, our system still outperforms existing baselines in spatial-aware ASR tasks, demonstrating its robustness and adaptability to different acoustic conditions. 
Other than speech, we also test the performance on audio dataset ESC-50~\cite{piczak2015esc}, also shows robust performance as speech.

\begin{table}[h]
    \centering
    \caption{Performance on Different Datasets: We evaluate our model across several datasets, reporting MAE and Median Error for DoA estimation.}
    \begin{tabular}{lcc}
        \toprule
        \textbf{Dataset} & \textbf{MAE (°) $\downarrow$} & \textbf{Median Error (°) $\downarrow$} \\
        \midrule
        LibriSpeech (speech) & \textbf{25.72} & \textbf{13.00} \\
        Common Voice (speech) & 53.34 & 32.11 \\
        VoxCeleb (speech) & 41.22 & 25.80 \\
        LJ Speech (speech) & 80.21 & 34.73 \\
        ESC-50 (audio) & 49.99 & 24.65 \\
        \bottomrule
    \end{tabular}
    \label{tab:dataset_performance}
\end{table}

\section{3D UMAP Visualization of Spatial Embeddings}\label{app:umap}
\begin{figure}[htbp]
\begin{center}
\vspace{-0.15in}
\includegraphics[width=7.0in]{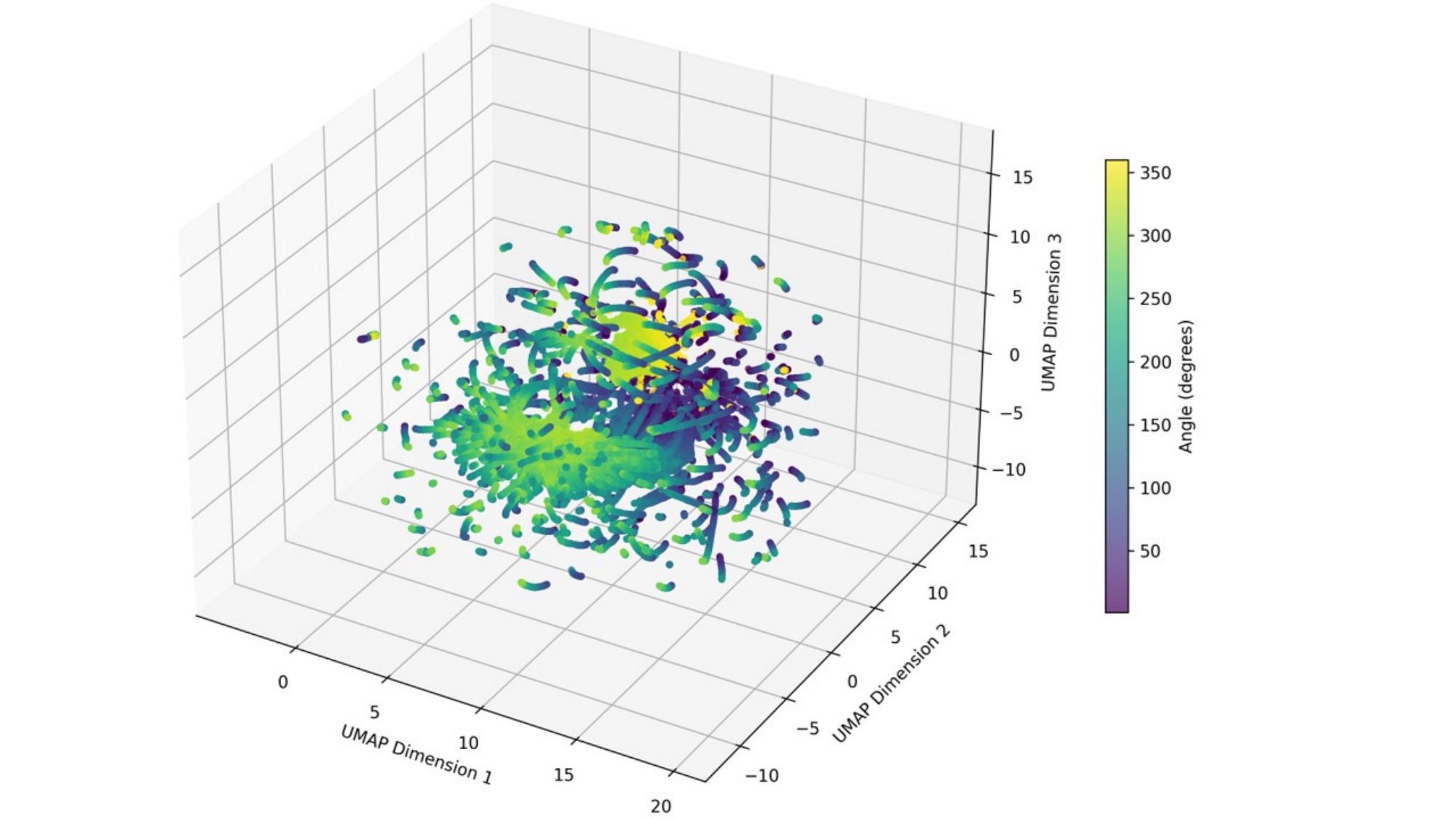}
\vspace{-0.15in}
\caption{\small{3D UMAP visualization of spatial embeddings generated by the DoA encoder.}}
\label{fig:umap}
\end{center}
\end{figure}

To evaluate the spatial embeddings derived from the DoA encoder, we utilized UMAP (Uniform Manifold Approximation and Projection)\cite{mcinnes2018umap} to visualize their structure in a reduced-dimensional space. Figure~\ref{fig:umap} showcases the 3D UMAP visualizations, where the embeddings capture spatial cues from all angles, effectively encoding directional information. The clustering and angular organization of the points demonstrate the encoder's capacity to represent spatial diversity and preserve the underlying spatial structure of the speech signals. This provides valuable insights into the model's ability to generalize spatial features across multiple sound sources.

\section{Cumulative Distributive Analysis for different speech sources for SING}\label{app:cdf}

\begin{figure*}[!htbp]
\centering

\begin{minipage}{0.49\textwidth}
    \centering
    \includegraphics[width=\textwidth]{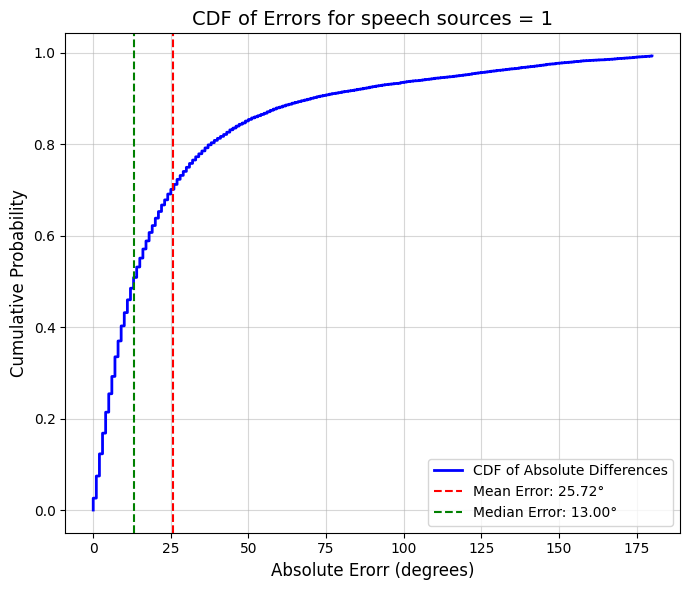} 
    \caption{CDF for 1 DoA}
    \label{fig:plot1}
\end{minipage}
\hfill
\begin{minipage}{0.49\textwidth}
    \centering
    \includegraphics[width=\textwidth]{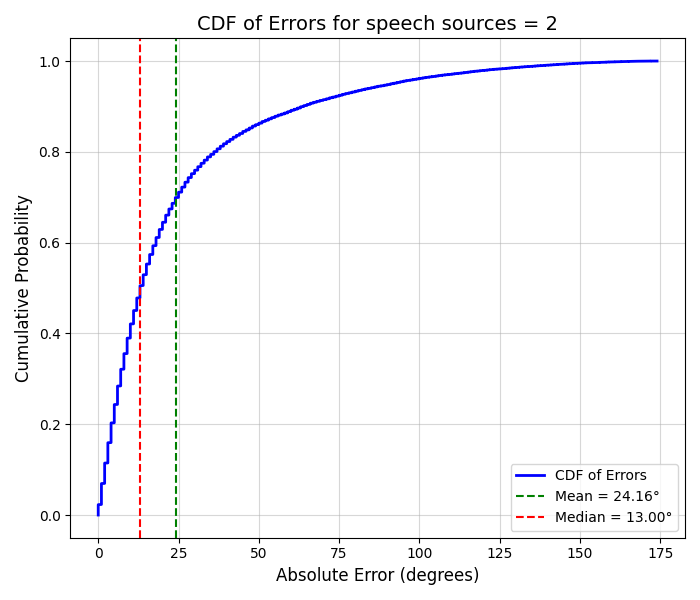} 
    \caption{CDF for 2 DoAs}
    \label{fig:plot2}
\end{minipage}

\begin{minipage}{0.49\textwidth}
    \centering
    \includegraphics[width=\textwidth]{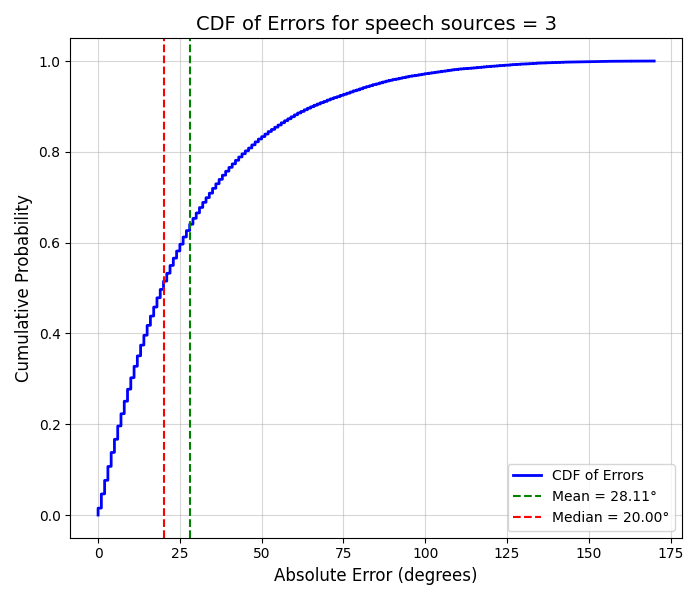} 
    \caption{CDF for 3 DoAs}
    \label{fig:plot3}
\end{minipage}
\hfill
\begin{minipage}{0.49\textwidth}
    \centering
    \includegraphics[width=\textwidth]{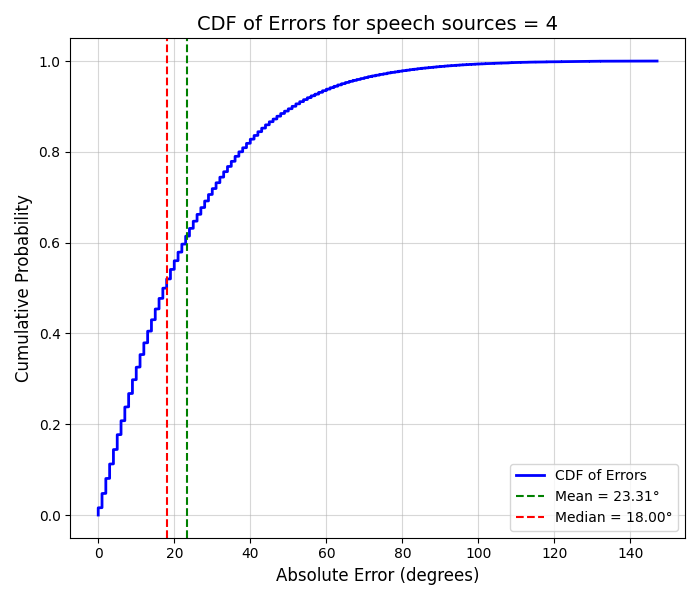} 
    \caption{CDF for 4 DoAs}
    \label{fig:plot4}
\end{minipage}

\begin{minipage}{0.49\textwidth}
    \centering
    \includegraphics[width=\textwidth]{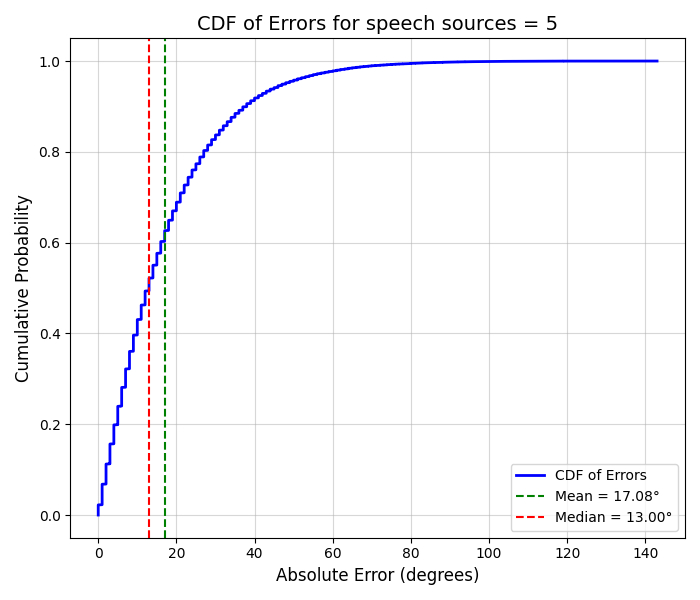} 
    \caption{CDF for 5 DoAs}
    \label{fig:plot5}
\end{minipage}
\hfill
\begin{minipage}{0.49\textwidth}
    \centering
    \includegraphics[width=\textwidth]{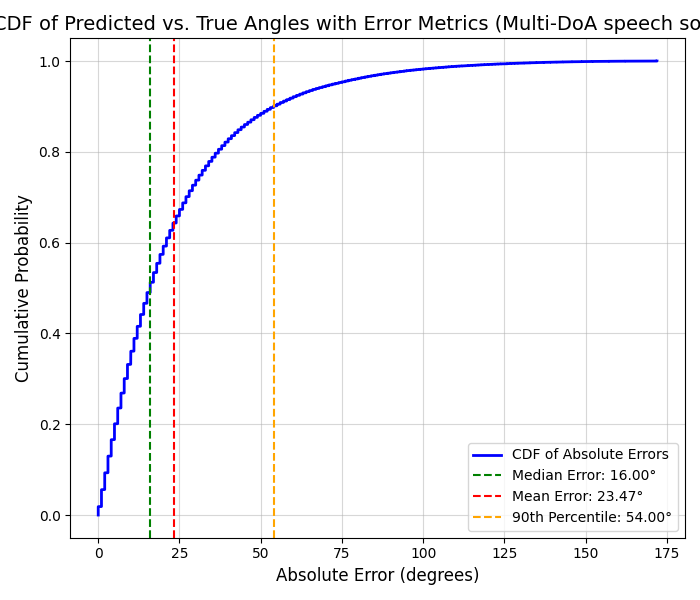} 
    \caption{CDF for all speech sources}
    \label{fig:plot6}
\end{minipage}

\caption{CDF plots of absolute errors for different numbers of speech sources (DoAs). Mean and median errors are highlighted for comparison.}
\label{fig:appendix_cdf_plots}
\end{figure*}

We focus on analyzing the mean and median errors across varying conditions, illustrating our findings through comprehensive plots for clarity. Figure~\ref{fig:appendix_cdf_plots} shows the cumulative distribution function (CDF) plots for different speech sources.






\section{Prompt Details}\label{app:prompts}

To evaluate the capabilities of our Spatial-Aware ASR system, we design various prompts corresponding to different speech processing tasks. These prompts facilitate direction estimation, speaker counting, and transcription summarization, enabling a more comprehensive understanding of spatially distributed speech signals. 

\begin{table*}[h!]
\centering
\begin{tabular}{@{}p{0.23\textwidth}p{0.23\textwidth}p{0.23\textwidth}p{0.23\textwidth}@{}}
\toprule
\textbf{Type} & \textbf{System Prompts} & \textbf{Prompts} & \textbf{Response Templates} \\ \midrule
ASR (Whisper) & You are an assistant that provides the summarization of the speech. Respond with the the correct summarization. & What is the summarization of the speech? & Mr. Swift's eyes twinkled. \\ \midrule
Spatial ASR (DoA + Whisper) & You are an assistant that provides the direction of the speech in degrees and a summarization of the speech. Respond with the exact angle in degrees and the correct summarization. & What is the direction of speech in degrees and summarization of the speech? & The speaker is speaking approximately at angle 217 degrees. The speech says: A low, deep moan broke from him.  \\ \midrule
DoA Detection & You are an assistant that identifies the direction of sound. Respond with the exact angle in degrees. & What is the direction of the speech source? & The speech is coming from 45 degrees. \\ \midrule
Number of Speakers & You are an assistant that identifies how many people are speaking at the same time. Respond with the exact numbers. & How many people are speaking? & There are 3 speech sources. \\ \midrule
Multi-DoA Detection & You are an assistant that identifies how many people are speaking at the same time and the directions of speech. Respond with the exact angle in degrees. & How many people are speaking? What are the directions of the sound sources? & There are 2 speech sources. Speech source 1’s degree of arrival is 39.0 degrees, and speech source
2’s degree of arrival is 101.0 degrees. \\ 

\bottomrule
\end{tabular}
\caption{Overview of different prompt types and their corresponding response templates in the Spatial-Aware ASR system. The system is capable of detecting speech direction, counting the number of speakers, and summarizing transcriptions.}
\label{tab:response_table}
\end{table*}

\section{Performance of Number of Speakers Estimation}\label{app:num-of-speaker-encoder}
\begin{figure}[htbp]
\begin{center}
\vspace{-0.1in}
\includegraphics[width=3.7in]{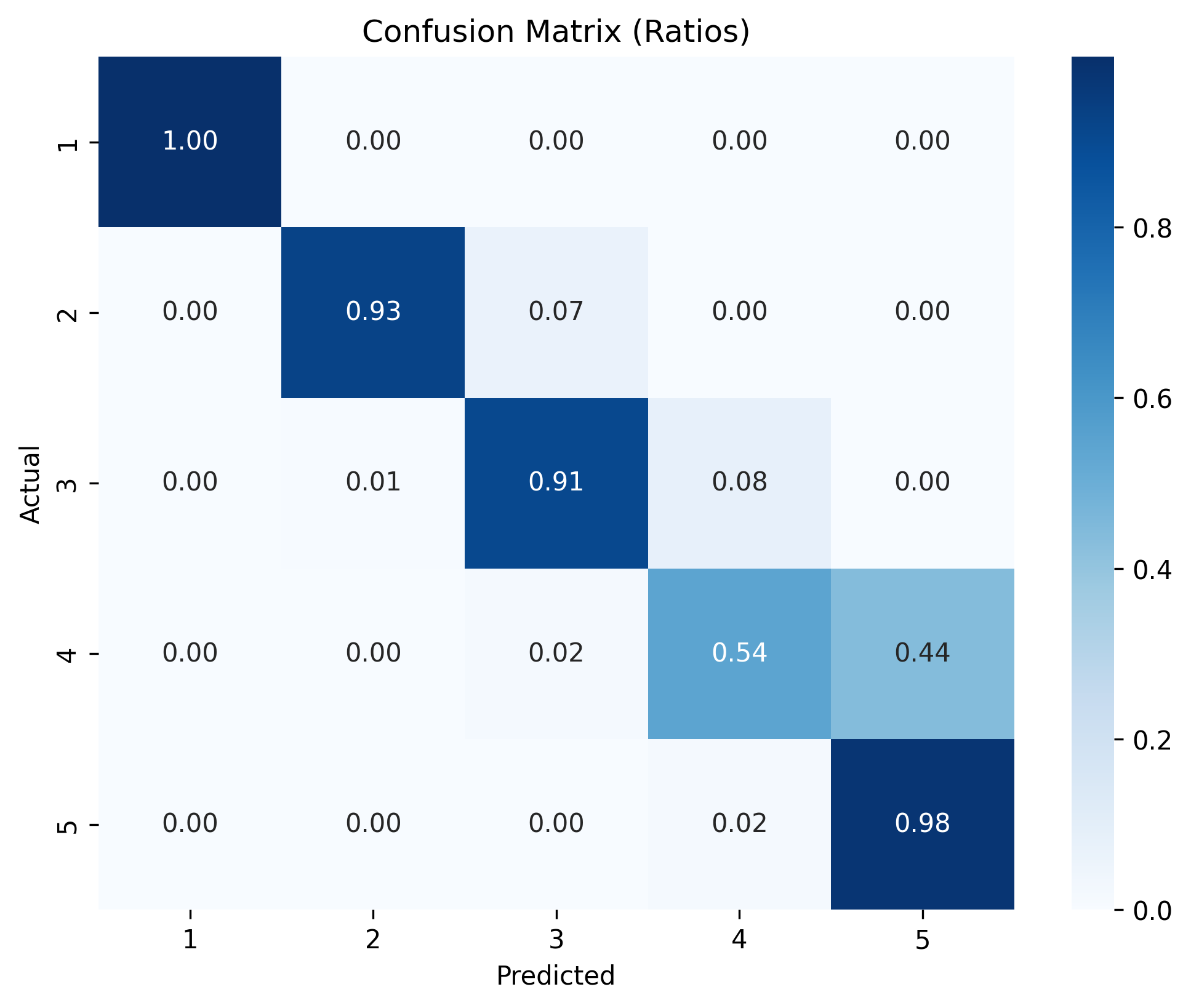}
\vspace{-0.2in}
\caption{\small{Confusion matrix for number of speakers estimation.}}
\vspace{-0.1in}
\label{fig:cross_correlation}
\end{center}
\end{figure}

We show the performance of number of speakers estimation from the num of speaker encoder. 
Figure~\ref{fig:cross_correlation} shows the normalized confusion matrix for the number of speakers' estimation task, represented as ratios. Each cell indicates the proportion of predictions for a given actual class, normalized by the total number of samples in that class. The diagonal values highlight the model's accuracy in correctly identifying the number of speakers, with values close to 1 indicating high precision. Off-diagonal values represent misclassifications, demonstrating the model's tendency to confuse certain classes. For example, the model exhibits strong performance for classes 1, 2, 3, and 5, while class 4 shows more confusion with class 5, reflecting the challenges in differentiating between these scenarios.

\section{Spectrogram of Recordings from Directions}\label{app:spectrogram}
\begin{figure*}[htbp]
\begin{center}
\includegraphics[width=6.9in]{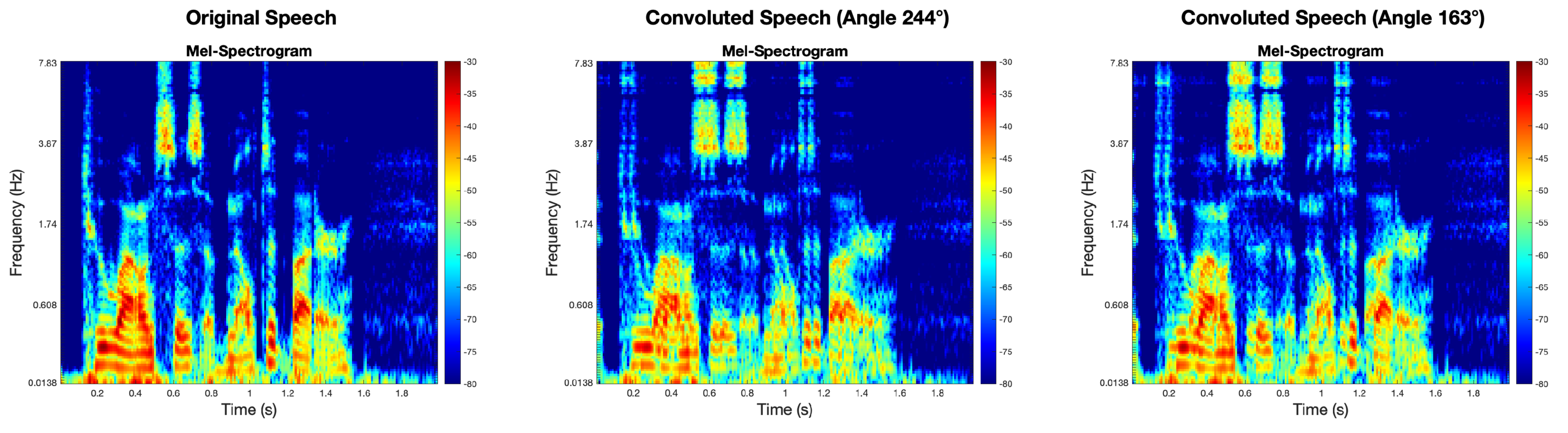}
\caption{\small{Comparison between original speech and speech captured from directions.}}
\label{fig:speech_spectrogram}
\end{center}
\end{figure*}

Figure~\ref{fig:speech_spectrogram} shows the difference between original speech and speech signals coming from angles, captured by the microphone built in microstructure.  
The differences in the spectrograms are evident in the amplitude and frequency distributions, which vary significantly depending on the Direction of Arrival (DoA). For example, at angles 244° and 163°, the spectrograms display distinct patterns in the intensity and distribution of energy across frequencies compared to the original speech. These variations are introduced by the microstructure's unique modulation of sound waves, which embeds direction-specific characteristics into the captured audio.

\end{document}